\documentclass[reqno]{article}

\usepackage{graphicx}
\usepackage{amsmath,amsthm,amssymb}

\chardef\bslash=`\\ 

\theoremstyle{definition}

\theoremstyle{remark}

\newcommand{\eval}[2][\right]{\relax
  \ifx#1\right\relax \left.\fi#2#1\rvert}

\begin{document}
\title{\bf{Periodic self-similar wave maps coupled to gravity}}

\author{Piotr Bizo\'n\footnotemark[1]{}\,\;, Sebastian Szybka\footnotemark[1]{}\,\;, and
  Arthur Wasserman\footnotemark[2]{}\\
  \footnotemark[1]{} \small{\textit{Institute of Physics,
   Jagellonian University, Krak\'ow, Poland}}\\
   \footnotemark[2]{} \small{\textit{Department of Mathematics,
   University of Michigan, Ann Arbor, Michigan}}}
\maketitle
\begin{abstract}
\noindent We continue our studies of spherically symmetric
self-similar solutions in the $SU(2)$ sigma model coupled to
gravity. Using mixed numerical and analytical methods we show
existence of an unstable periodic solution lying at the boundary
between the basins of two generic attractors.

\end{abstract}

\section{Introduction}
This is the third paper in a series aimed at understanding the
structure of self-similar spherically symmetric wave maps coupled
to gravity. In the first two papers~\cite{p1,p2} we showed that
for small values of the coupling constant there exists a countable
family of solutions that are analytic below the Cauchy horizon of
the central singularity. In this paper we wish to elaborate on the
analysis of a periodic self-similar solution whose existence was
only briefly mentioned in~\cite{p1}. We tried to make this paper
self-contained mathematically but we refer the reader to~\cite{p1}
for the discussion of the physical background of the problem and
to~\cite{husa1} for more on the role of self-similar solutions in
gravitational collapse.
\section{Setup}
 We showed in~\cite{p1} that under the assumptions of spherical symmetry,
 equivariance and self-similarity the Einstein equations coupled to the $SU(2)$
  sigma field  reduce to
  the following
system of autonomous ordinary differential equations for three
functions $W(x), A(x)$, and $F(x)$
\begin{eqnarray}\label{main}
W'&=&-1+\alpha(1-W^{2})F'^{2},\\
A'&=&-2\alpha A W F'^{2},\\
(AF')'&=&\frac{\sin(2F)}{W^{2}-1}.
\end{eqnarray}
subject to the constraint
\begin{equation}\label{con}
1-A-2\alpha \sin^2\!{F}+\alpha AF'^{2}(W^{2}-1)=0.
\end{equation}
Physically the functions $W$ and $A$ parametrize the metric, the
function $F$ parametrizes the $SU(2)$ sigma field and $\alpha$ is
the dimensionless coupling constant ($\alpha=0$ corresponds to no
gravity). We are interested in solutions of equations (1)-(4)
starting at say $x=0$ with the following initial conditions (as
explained in~\cite{p1} these conditions ensure regularity of
solutions at the past light cone of the singularity)
\begin{equation}\label{initial}
 \quad F(0)=\frac{\pi}{2}, \quad
F'(0)=b, \quad W(0)=1, \quad A(0)=1-2\alpha,
\end{equation}
where $b$ is a free parameter (since the system has reflection
symmetry $F\rightarrow -F$ we may take $b>0$ without loss of
generality). The value $A(0)$ follows from the constraint (4).
 In
what follows we shall refer to solutions of equations (1)-(4)
satisfying the initial conditions (\ref{initial}) as $b$-orbits.
We showed in~\cite{p2} that for $\alpha <\frac{1}{2}$, $b$-orbits
exist locally and are analytic in $b$ and $x$.

 It follows
immediately from (1)-(4) that a $b$-orbit can be continued as long
as $|W|<1$ (since then $A$ is bounded away from zero). However, if
$W$ hits $\pm 1$ at some $x$, then the solution becomes singular.
Now we shall show that generic $b$-orbits become singular in
finite time. Throughout the paper we assume that $\alpha
<\frac{1}{2}$.

We first show that if $b$ is small, then $W$ tends to -1 for some
finite $x_{A}$. To see this, let $f=(F-\pi/2)/b$. Then, in the
limit $b \rightarrow 0$, equations (1)-(3) reduce to
\begin{eqnarray}
W' &=& -1 +\alpha (1-W^2) b^2 {f'}^2 \rightarrow -1,\\
A'&=& -2 \alpha A W\: b^2 {f'}^2 \rightarrow 0, \\
(A f')'& =& -\frac{\sin(2 b f)}{b (W^2-1)} \rightarrow -\frac{2
f}{W^2-1},
\end{eqnarray}
with the initial conditions
\begin{equation}
f(0)=0, \quad f'(0)=1,\quad W(0)= 1, \quad A(0)=1-2\alpha.
\end{equation}
The limiting  equations (6) and (7) are solved by $W=1-x$ and
$A=1-2\alpha$. Substituting these solutions into (8) we get the
equation
\begin{equation}
(1-2 \alpha) f''+ \frac{2 f}{x (x-2)} =0,
\end{equation}
whose solution (which can be obtained in closed form) behaves as
$f(x) \sim (1- 2 \alpha) + (2-x) \ln(2-x)$ for $x \rightarrow 2$.
Thus,  the term $(1-W^2) {f'}^2$ in (6) stays bounded so,
  by uniform convergence
 on compact
intervals, the solutions with sufficiently small $b$ will tend to
$W=-1$.

Next, we show that if $b$ is large then $W$ tends to $+1$ and
$A\rightarrow 0$ for some finite $x_{B}$. This time we define the
variables
\begin{equation}
\xi=b^2 x, \quad h(\xi)=b^2 (1-W(x)), \quad s(\xi)=
 b(F(x)-\frac{\pi}{2}).
\end{equation}
Then, in the limit $b\rightarrow \infty$, equations (1)-(3) reduce
to (where now prime is $d/d\xi$)
\begin{eqnarray}
h' &=& 1 -\alpha h \left(2-\frac{h}{b^2}\right) {s'}^2 \rightarrow
1-2\alpha h {s'}^2,\\
A'&=& -2 \alpha A \left(1-\frac{h}{b^2}\right) {s'}^2  \rightarrow
-2\alpha A {s'}^2,
\\
(A s')'& =& \frac{\sin(2 s/b)}{b h (2-h/b^2)} \rightarrow 0,
\end{eqnarray}
with the initial conditions
\begin{equation}
h(0)=0, \quad A(0)=1-2\alpha, \quad s(0)=0, \quad s'(0)=1.
\end{equation}
It follows from (14) and (15) that, in the limit $b \rightarrow
\infty$, $A s'=1-2 \alpha$. Plugging this into equations (12) and
(13), and using (15), we get the limiting solution
\begin{equation}
A(\xi)=(1-2 \alpha) \sqrt{1-4 \alpha \xi} \quad , \quad h(\xi)=
\frac{1}{2\alpha} \sqrt{1-4 \alpha \xi}\: (1-\sqrt{1-4 \alpha
\xi}\:).
\end{equation}
This solution becomes singular at $\xi=1/4\alpha$. By uniform
convergence on compact intervals, we conclude
 that for solutions
of equations (1)-(4) with large $b$ (and nonzero $\alpha$), the
function
 $W(x)$ attains a
minimum and then tends to $1$ at some $x \rightarrow x_B$, while
the function $A(x)$ drops to zero at $x_B$.

 To summarize,
$b$-orbits tend in finite time to $W=-1$ if $b$ is small, and to
$W=+1$ if $b$ is large. In what follows, we shall refer to these
two kinds of solutions as to type $A$ and type $B$ orbits,
respectively. We show next that the sets of type $A$ and type $B$
orbits are open. \vskip 0.12cm \noindent \textbf{Lemma 1.} If
$W(x)>0$ and $A(x)<1/2-\alpha$ for some $x>0$ then the orbit is of
type B, i.e., there is a finite $x_0$ such that
$\lim\limits_{x\rightarrow x_0} W(x)=1$. Moreover,
$\lim\limits_{x\rightarrow x_0} A(x)=0$.
 \vskip
0.12cm \noindent \emph{Proof:}
 Substituting equation (4) in (1) we get
\begin{equation}
W'=-2+\frac{1-2\alpha \sin^2{F}}{A}>-2+\frac{1-2\alpha}{A}.
\end{equation}
Thus, if $A(x)<1/2-\alpha$ then $W'(x)>0$ so if $W(x)>0$ then $W$
remains positive. But then by equation (2) $A$ decreases which
implies by (17) that $W'$ remains positive (bounded away from zero
in fact) and hence $W$ must hit $+1$ in finite time. To prove the
second part of the lemma, note that by equations (1) and (2) we
have (using the abbreviation $V=1-W^2$)
\begin{equation}\label{V}
    \left(\frac{V}{A}\right)'= \frac{2 W}{A}.
\end{equation}
Assume that $A(x_0)>0$. Then $(V/A)(x_0)=0$  and since
$(V/A)(x)>0$ for $x<x_0$ we get a contradiction. Hence $A(x_0)=0$.
\vskip 0.12cm \noindent \textbf{Corollary.} Type B orbits are
open. \vskip 0.12cm \noindent \emph{Proof:} If the $b_0$-orbit is
of type B then by lemma~1, $A(x,b_0)<1/2-\alpha$ and $W(x,b_0)>0$
for some $x>0$. Hence for nearby $b$ we also have
$A(x,b)<1/2-\alpha$ and $W(x,b)>0$ and thus, again by lemma~1, the
$b$-orbit is of type B. \vskip 0.12cm \noindent\textbf{Proposition
1.} Type A orbits are open. \vskip 0.12cm \noindent\emph{Proof:}
First, note that if the orbit is of type A and $W(x)\geq 0$ then
$A(x)\geq 1/2-\alpha$ (since otherwise the orbit would be of type
B by lemma~1). But for $W<0$, by equation (2) $A'>0$, hence
$A(x)>A(x_0)\geq 1/2-\alpha$ for $x\geq x_0$ where $x_0$ is the
point at which $W(x_0)=0$. Thus, $A>1-2\alpha$ for type A orbits.
Now, let $b_0$-orbit be of type A and consider a nearby $b$-orbit.
By continuity, there is a point $x_1$ such that $W(x_1,b)$ is
close to $-1$, $W(x_1,b)<0$, and $A(x_1,b)$ is greater than, say,
$1/2-\alpha$.
 First we show that
such orbits have $W'(x,b)<0$ for all $x>x_1$. To see this, notice
that from equations (1)-(3)
$$W''=-2\alpha F'A^{-1}\left(\alpha A W{F'}^{3}(W^{2}-1)-W A F'+\sin(2F)\right),$$
hence at the first zero of $W'(x,b)$ after $x_1$ we have
\begin{equation}\label{wbis}
W''\Big{\vert}_{W'=0}=\frac{1}{A V}(4A W\pm 2\sqrt{\alpha
V}\sin{2F}).
\end{equation}
The numerator is negative because $A>1/2-\alpha$ and $W$ is close
to $-1$ while the denominator is always positive, hence $W''<0$
which is a contradiction. Thus, $W'(x,b)<0$ and
$\lim\limits_{x\rightarrow x_1} W(x,b)$ exists (if the orbit stays
in the region). Now we show that $W(x_2,b)=-1$ for any such orbit
for some $x_2>x_1$.
 To prove this assume  that
$\lim\limits_{x\rightarrow\infty}W(x)=\bar{W}\geq -1$. Integrating
equation (1) we get
\begin{equation}\label{a}
\int_{x_1}^{x} W' dx = x_1-x + \alpha \int_{x_1}^{x} V F'^2 dx,
\end{equation}
which gives a contradiction as $x\rightarrow \infty$ because the
last integral in (\ref{a}) is finite by equation (2) (remember
that $A$ is bounded from below).

We now know that both type A and type B orbits are open so there
must be orbits that are not type A or type B, that is, orbits that
stay in $W^2<1$ for all $x$. Call these type C orbits. We note
that type C orbits are defined for all $x\geq 0$ since
$A(x)>1/2-\alpha$ for all $x$. \vskip 0.11cm \noindent
\textbf{Proposition 2.} For any type C orbit we have
\begin{equation}
\liminf W(x) \leq 0 \quad \mbox{and} \quad \limsup W(x)\geq 0.
\end{equation}
\vskip 0.11cm \noindent\emph{Proof:} Suppose that there is an
$x_1$ such that $W(x)\leq -L<0$ for $x>x_1$ (this is equivalent to
$\limsup W(x)< 0$). Then from equation (~\ref{V})
$$
\left(\frac{V}{A}\right)'= \frac{2 W}{A} \leq \frac{-L}{A}\leq-L
$$
for $x>x_1$ and hence $(V/A)<0$ for some $x_2>x_1$ which is a
contradiction since $(V/A)\geq 0$ for all $x$.

Similarly, suppose that there is an $x_1$ such that $W(x)\geq L>0$
for $x>x_1$ (this is equivalent to $\liminf W(x)> 0$). From
lemma~1 we have that $A(x)>1/2-\alpha$ for $x>x_1$. Thus
$$
\left(\frac{V}{A}\right)'= \frac{2 W}{A} \geq 2 L
$$
for $x>x_1$ which is a contradiction since by lemma~1 $(V/A)\leq
2/(1-2\alpha)$ for all $x$.

From Proposition 2 we see that type C orbits must oscillate at
infinity about $W=0$ (unless $\lim W=0$).
 \vskip 0.1cm Once we know that type
$C$ orbits exist we turn to their numerical construction.
 Numerics indicates that the structure
of type $C$ orbits is rather complicated for large
$\alpha$~\cite{seb}. In this paper we restrict our attention to
small values $\alpha\le 0.42$, where the structure is simple.
Namely for each given $\alpha$ there is a single critical value
$b^{*}(\alpha)$ such that $b$-orbits tend to the attractor $A$
(resp. $B$) if $b<b^{*}$ (resp. $b>b^{*}$) and the $b^{*}$-orbit
is of type $C$. In other words, the $b^{*}$-orbit is a separatrix
lying between two generic attractors $A$ and $B$. In the next
section we give numerical and analytical arguments that the
$b^{*}$-orbit is asymptotically periodic.
\section{Numerical solution}
A straightforward way to determine the critical value $b^{*}$ is
to take two values $b_{A}$ and $b_{B}$ leading to attractors $A$
and $B$ respectively and then fine-tune to $b^{*}$ by bisection.
This procedure yields a pair of $b$ that are within a distance
$\epsilon$ from $b^{*}$ (where $\epsilon$ is limited by machine
precision). Such marginally critical $b$-orbits exhibit a
transient periodic behavior before eventually escaping toward
$W=\pm 1$ (see Figures 1 and 2).
\begin{figure}[!h]
\centering
\includegraphics[width=0.72\textwidth]{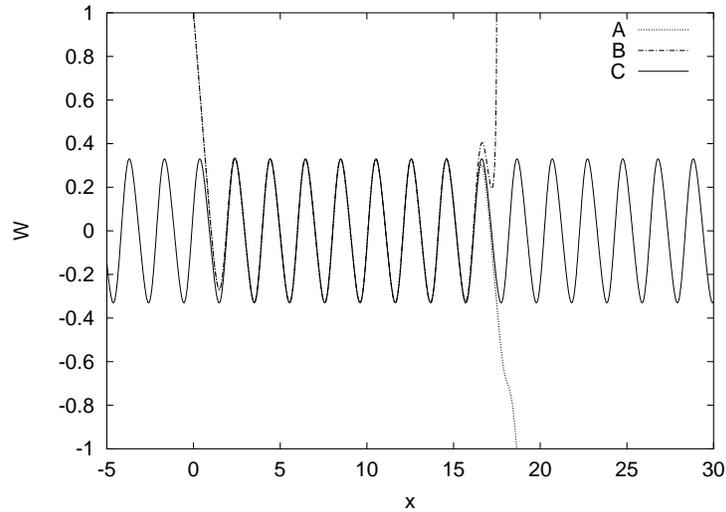}
\caption{\small{The function $W(x)$ for two marginally critical
$b$-orbits for $\alpha=0.38$: the type A solution with
$b=b^*-\epsilon$ (dotted line) and the type B solution with
$b=b^*+\epsilon$ (dashed-dotted line), where $\epsilon=10^{-17}$.
Superimposed (solid line) is the periodic solution constructed by
the straddle-orbit method.}}
\end{figure}
 \begin{figure}[!h]
\centering
\includegraphics[width=0.7\textwidth]{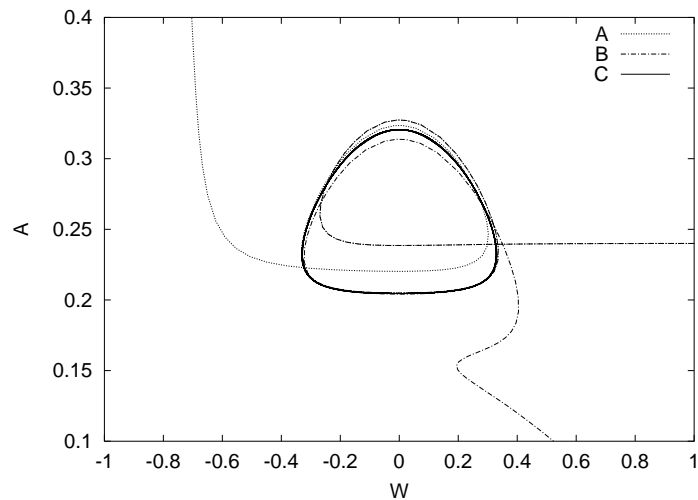}
\caption{\small{The projection on the $(A,W)$ plane of the same
solution as in Fig.~1. The periodic solution is seen as the
unstable limit cycle.}}
\end{figure}

\noindent This suggests that the system has an unstable periodic
solution and the $b^{*}$-orbit belongs to its basin of attraction.
In other words, the value $b^{*}$ corresponds to the intersection
of the line of initial data ($W=1$, $F=0$, $F'=b$) with the
2-dimensional stable manifold of the periodic solution. In fact,
if we take any two points $P_{A}$ and $P_{B}$ in the phase space
which lead to attractors $A$ and $B$, respectively, and perform
bisection we obtain the same asymptotically periodic solution.
This indicates that the stable manifold of the periodic solution
is the  boundary between the basins of attractors $A$ and $B$.

 Since the
periodic solution is unstable and numerically it is impossible to
set initial conditions exactly on the stable manifold, we cannot
obtain too many cycles of the periodic solution. Although in our
case this is not a serious difficulty because the positive
Lyapunov exponents are not large (see Figure 7 below), we would
like to remark in passing that using so called straddle-orbit
method due to Battelino et al.~\cite{straddle} one can pursue the
unstable periodic orbit in principle forever. This procedure,
which can be viewed as a series of bisections, goes as follows. At
initial time we choose two points $P_{A}(x=0)$ and $P_{B}(x=0)$
which lead to different attractors $A$ and $B$ and perform
bisection until the distance between the iterates $P_{A}(0)$ and
$P_{B}(0)$ is less a prescribed $\delta$. Next we integrate the
equations numerically starting from the current $P_{A}(0)$ and
$P_{B}(0)$ until the distance between the trajectories exceeds
$\delta$. When this happens at some time $x$ we stop the
integration, assign the points $P_{A}(x)$ and $P_{B}(x)$ as
current representatives and repeat the bisection. Iterating this
procedure one can progressively construct a trajectory staying
within a distance $\delta$ from the codimension one stable
manifold. The numerical solutions obtained by this method are
shown in Figures 3, 4, and 5.
 \begin{figure}[!h]
\centering
\includegraphics[width=\textwidth]{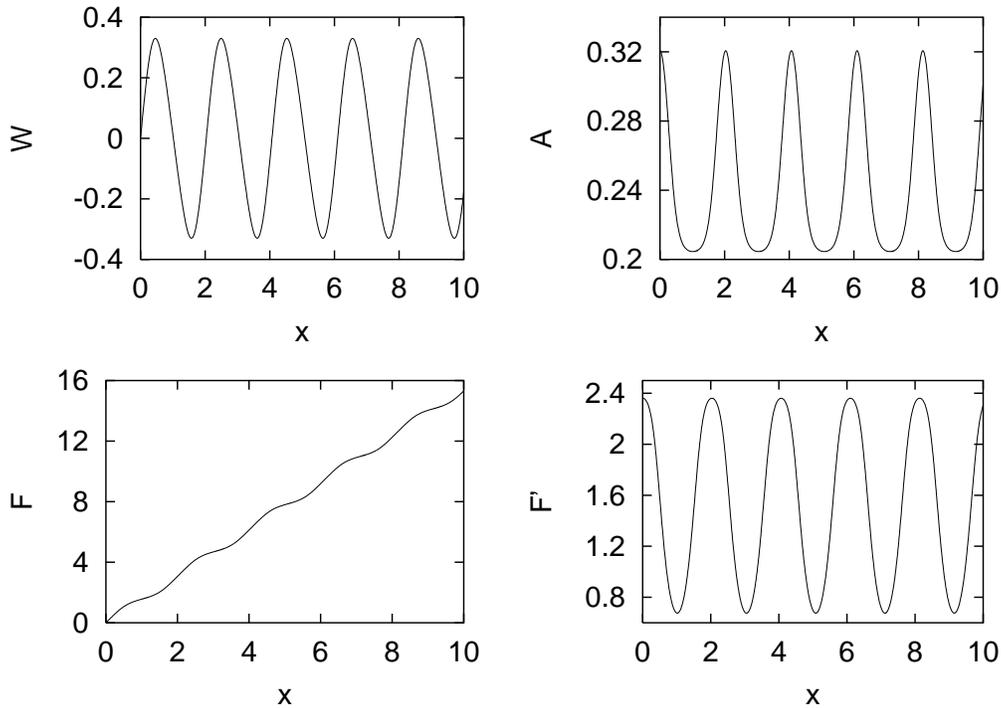}
\caption{\small{The profiles of the periodic solution for
$\alpha=0.38$.}}
\end{figure}
\newpage
 \begin{figure}[!h]
\centering
\includegraphics[width=0.65\textwidth]{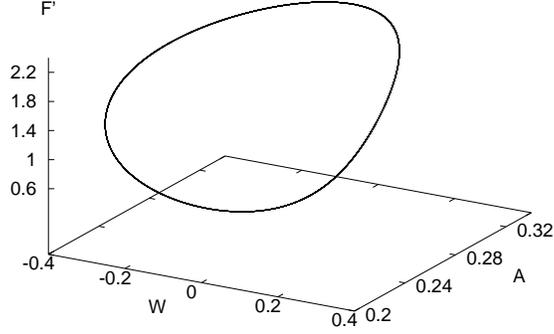}
\caption{\small{The phase portrait of the periodic solution for
$\alpha=0.38$.}}
\end{figure}
 \begin{figure}[!h]
\centering
\includegraphics[width=0.65\textwidth]{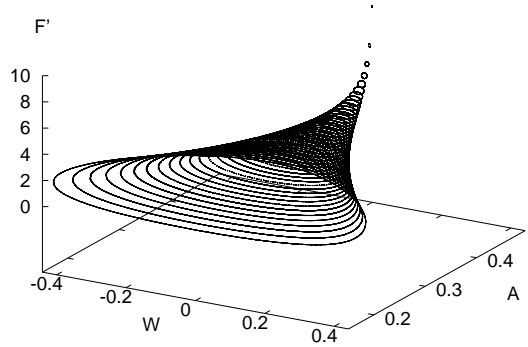}
\caption{\small{The phase portraits of periodic solutions for
different values of the coupling constant $\alpha$ ranging from
$0.01$ to $0.42$. As $\alpha\rightarrow 0$ the loop shrinks to
zero and $F'\rightarrow \infty$.}}
\end{figure}

\noindent Using the fact that non-critical $b$-orbits become
singular in finite time we can easily compute the positive
Lyapunov exponent $\lambda$ of the periodic solution. To this end
consider a marginally critical $b$-orbit with $b=b^*-\epsilon$.
Such an orbit approaches the periodic solution, stays close to it
for some time and eventually escapes along the unstable manifold
to crash at a point $x_A$ where $W(x_A)=-1$. Therefore we can
write
\begin{equation}\label{scaling}
x_A=x_{approach} + x_{periodic} + x_{escape},
\end{equation}
where $x_{approach}, x_{periodic}$, and $x_{escape}$ denote the
lengths of respective intervals of evolution (we say that the
solution "escapes" if its distance from the periodic attractor
exceeds a prescribed value). During the periodic interval the
distance between the $b$-orbit with $b=b^*-\epsilon$ and the
periodic solution grows at the rate proportional to $\epsilon
\exp(\lambda x)$, hence $x_{periodic} \sim (-1/\lambda)
\ln{\epsilon}$. This implies that the number of cycles $n$ during
this interval behaves as $n \sim (-1/\lambda T) \ln{\epsilon}$,
where $T$ is the period of the periodic solution. The length of
the escape interval does not depend on the number of cycles but
only on the phase of a cycle at which the escape from the periodic
solution takes place, hence $x_{escape} \sim f(\ln{\epsilon})$,
where $f$ is a periodic function with period $\lambda T$.
Summarizing, we have
\begin{equation}\label{wiggles}
x_A \approx -\frac{1}{\lambda} \ln{\epsilon} + f(\ln{\epsilon}) +
const.
\end{equation}
The numerical verification of this formula is shown in Figure 6.
Using (\ref{wiggles}) we calculated the dependence of $\lambda$ on
the coupling constant $\alpha$ - the result is shown in Figure 7.
 \begin{figure}[!h]
\centering
\includegraphics[width=0.6\textwidth]{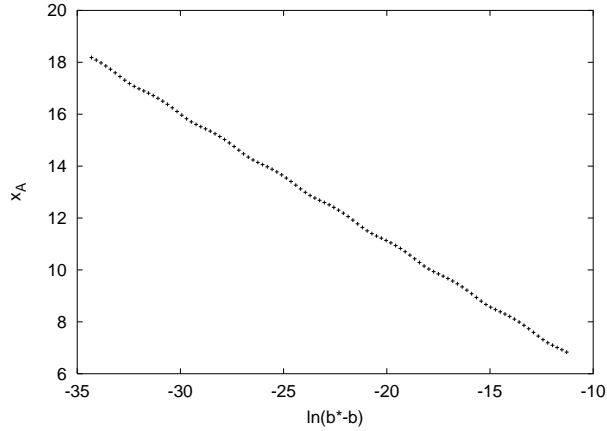}
\caption{\small{For $\alpha=0.2$, the locus of the point of crash
$x_A$ is plotted as the function of the logarithmic distance from
the critical value $\ln{\epsilon}$. The fit to the formula (23)
gives $\lambda=2.029$. The period of the wiggles, corresponding to
the function $f(\ln{\epsilon})$, is equal to $2.887$ in agreement
with the predicted value $\lambda T$ (where $T=1.418$ was
calculated independently from equation (34)) }.}
\end{figure}
 \begin{figure}[!h]
\centering
\includegraphics[width=0.6\textwidth]{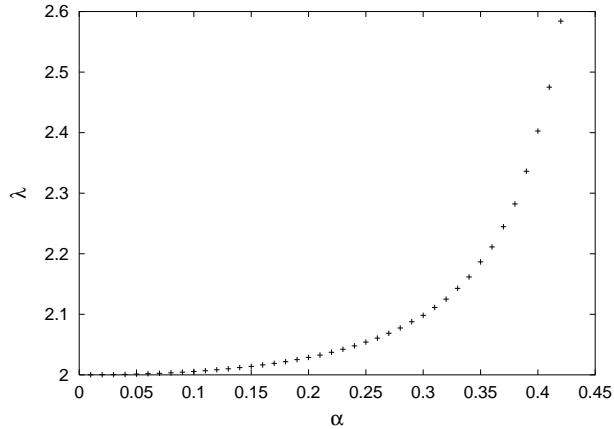}
\caption{\small{The positive Lyapunov exponent $\lambda$ of the
periodic solution as a function of the coupling constant
$\alpha$.}}
\end{figure}
\section{Perturbation series}
In order to construct periodic solutions we consider equations
(1)-(4) with initial conditions
\begin{equation}\label{initial2}
 \quad F(0)=0, \quad
F'(0)=c, \quad W(0)=0, \quad A(0)=(1+c^2)^{-1},
\end{equation}
where $c$ is a free parameter and the value $A(0)$ follows from
the constraint (4). We claim that for sufficiently small $\alpha$
there is a unique $c$ such that
 $F(T)=\pi,
F'(T)=F'(0), W(T)=W(0), A(T)=A(0)$ for some $T>0$. Since the
system is invariant under the  shift $F\rightarrow F+\pi$, we call
such solution periodic.  Now we shall construct the periodic
solution in a perturbative way using the Poincar\'e-Lindstedt
method \cite{poincare}.

We define the new variable $y=\frac{\omega x}{\sqrt{\alpha}}$
where $\omega$ is the unknown in advance frequency. We remark that
the rescaling of the independent variable by $\sqrt{\alpha}$ is
essential in order to have a well-defined limit for
$\alpha\rightarrow 0$ while the rescaling by $\omega$ is
introduced for convenience in order to have the fixed  period
$2\pi$. In terms of $y$ equations (1)-(3) transform to
$(\beta=\sqrt{\alpha})$
\begin{eqnarray}\label{main2}
\omega W'&=&\beta[-1+\omega^2(1-W^{2})F'^{2}],\\
A'&=&-2\omega\beta A W F'^{2},\\
{\omega}^{2}(AF')'&=&\beta^{2} \frac{\sin(2F)}{W^{2}-1}.
\end{eqnarray}
and the constraint (4) becomes
\begin{equation}\label{con2}
1-A-2\beta^{2} \sin^2{F}+{\omega^2} AF'^{2}(W^{2}-1)=0.
\end{equation}
We consider these equations on the interval $0\leq y\leq 2\pi$
with the  boundary conditions
\begin{equation}\label{periodic}
F(0)=0,\quad  F(2\pi)=\pi,\quad W(0)=0,\quad A(0)=A_{0},
\end{equation}
where the value of the constant $A_{0}$ follows from the
constraint (28). We seek solutions in the form of a power series
in $\beta$
\begin{equation}\label{series}
W(y,\beta)=\sum_{k=0}^{\infty} \beta^{k} W_{k}(y), \quad
A(y,\beta)=\sum_{k=0}^{\infty} \beta^{k} A_{k}(y), \quad
F(y,\beta)=\sum_{k=0}^{\infty} \beta^{k} F_{k}(y).
\end{equation}
The key idea of the Poincar\'e-Lindstedt method is to expand
the frequency in the power series
\begin{equation}\label{seriesfreq}
\omega(\beta)=\sum_{k=0}^{\infty} \beta^{k} \omega_{k}.
\end{equation}
and to solve for the coefficients $\omega_{k}$ by demanding that
the solution contains no secular terms. Thus, we substitute
(\ref{series}) and (\ref{seriesfreq}) into (25)-(28), group the
terms according to powers of $\beta$ and require that the
coefficients of each power of $\beta$ vanish separately. In the
lowest order $O(1)$ we get
\begin{equation}\label{o1}
W_{0}(y)=0, \quad
A_{0}(y)=\left(1+\frac{{\omega_{0}}^{2}}{4}\right)^{-1}, \quad
F_{0}(y)=\frac{y}{2},
\end{equation}
where $\omega_{0}$ is yet undetermined. In the next order we get
the equation $\omega_{0}W_{1}'=(-1+\omega_{0}^{2}/4)$, so to avoid
a secular term we need to have $\omega_{0}=2$. Then, all
$O(\beta)$ terms are zero and in the order $O(\beta^2)$ we get
\begin{equation}\label{o2}
W_2(y)=0, \quad A_{2}(y)=-\frac{1}{2},\quad
F_{2}(y)=\frac{1}{2}\sin(y).
\end{equation}
Iterating this procedure with the help of Mathematica we
calculated the perturbation series up to order $O(\beta^{23})$.
For example, up to  order $O(\beta^{8})$ we have
\begin{eqnarray}
\omega(\beta)&=&2-\frac{\beta^4}{2}+\frac{\beta^6}{2}-\frac{49}{32}\;\beta^8 + O(\beta^{10}),\\
W(y,\beta)&=&
   \sin(y)\;{\beta }^{3} +
     \frac{\sin(2y)}{4}\;
  {\beta }^{5} +
  \frac{25 \sin(y) -
       5 \sin(2y)+
       \sin(3y)}{16}\;{\beta}^7 + O(\beta^9),\\
A(y,\beta) &=& \frac{1}{2} - \frac{\beta^2 }{2} +
  \frac{-2 +
       4 \cos (y)}{8}\;{\beta }^4
  +
     \frac{-4 - 8 \cos(y) +
       5 \cos (2y)}{16}\;{\beta }^6 \nonumber \\
       &+& \frac{-60  + 204
      \cos (y) -
     102 \cos (2y) + 52\cos (3y)}{384} \;{\beta }^8 + O(\beta^{10}),\\
F(y,\beta)&=& \frac{y}{2} + \frac{\sin(y)}{2}\;\beta^2 +
{16}\sin(2y)\; \beta^4 + \frac{81\sin(y) - 21\sin(2y)
+ \sin(3y)}{96}\; \beta^6 \nonumber +\\
&+& \frac{1656\sin(y) + 900\sin(2y) - 616\sin(3y) +
 9\sin(4y)}{4608}\;\beta^8 + O(\beta^{10}).
\end{eqnarray}
 We recall  that the ``physical'' frequency is equal to
$\frac{\omega}{\beta}$ so it diverges as $\beta$ tends to zero
(while the amplitude of oscillations goes to zero). In this sense
the periodic solution is nonperturbative even though we
constructed it by the perturbation technique.

For small values of $\beta$ the perturbation expansion converges
quickly to the periodic solution constructed numerically (see
Figure 8).  As $\beta$ grows the convergence becomes slower and we
need to take many terms in the perturbation series to approximate
well the numerical solution (see Figure 9).  The fact that two
independent ways of constructing the periodic solution agree,
makes us feel confident that the periodic solution does in fact
exist.
 \begin{figure}[!h]
\centering
\includegraphics[width=0.8\textwidth]{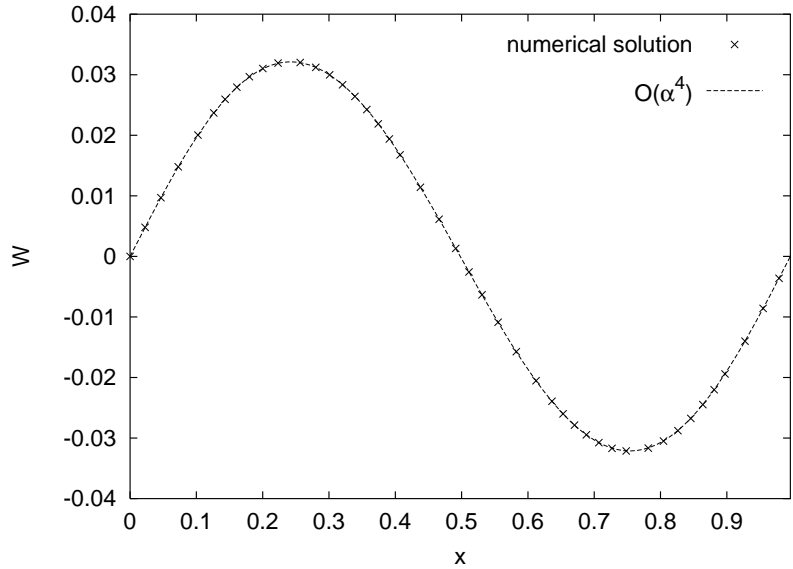}
\caption{\small{For $\alpha=0.1$ we plot the numerical periodic
solution and superimpose the perturbation series (35). Even at
this low order the agreement is very good.}}
\end{figure}
 \begin{figure}[!h]
\centering
\includegraphics[width=0.8\textwidth]{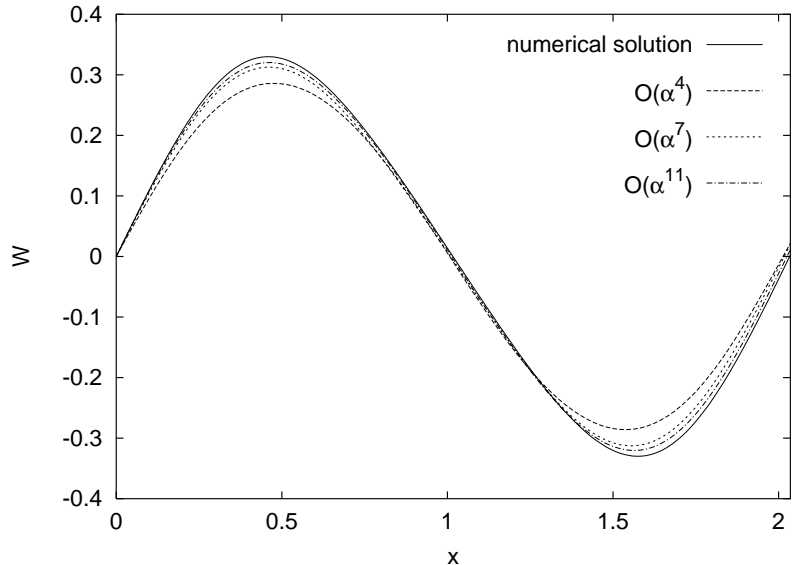}
\caption{\small{For $\alpha=0.38$ we plot the numerical periodic
solution and superimpose the perturbation series in different
orders. As the order increases the perturbation series slowly
approaches the numerical solution.}}
\end{figure}
\section{Final remarks}
We showed above that for small values of the coupling constant
$\alpha$, the critical $b^*(\alpha)$-orbit is asymptotically
periodic as $x\rightarrow \infty$. In the preceding papers [1,2]
we showed that for a generic value of $\alpha$, the
$b^*(\alpha)$-orbit evolved backwards in $x$ becomes singular as
$x\rightarrow -\infty$ (which corresponds to the singularity at
the center). However, there exist isolated values of $\alpha$
(called $\alpha_n$, $n=0,1,...$) for which the $b^*(\alpha)$-orbit
is regular as $x\rightarrow -\infty$. Combining this with the
result obtained above, we conclude that for a finite set of
isolated values $\alpha_n$ (satisfying $\alpha_n<0.42$) the
Einstein-wave map equations admit  self-similar solutions that are
regular at the center and asymptotically periodic outside the past
light cone.
\section*{Acknowledgement}
The research of PB and SS was supported in part by the KBN grant 2
P03B 006 23 and  the FWF grant P15738.

\end{document}